\documentclass{aa}
\usepackage{graphicx}
\usepackage{txfonts}
\begin{document}


\title{The ESO Nearby Abell Cluster Survey
       \thanks{Based on observations collected at the European Southern
               Observatory (La Silla, Chile)}}

\subtitle{VIII. Morphological and spectral classification of galaxies}
\author{T.~Thomas, P.~Katgert}
\institute{ Sterrewacht Leiden, The Netherlands}
\offprints{P.~Katgert}
\date{Received date; accepted date}
\markboth{The ESO Nearby Abell Cluster Survey VIII}{Morphological and
spectral classification of galaxies}

\abstract{
We determine the morphological types of 2295 galaxies from the ESO
Nearby Abell Cluster Survey (ENACS) from CCD images obtained with the
Dutch telescope on La Silla. A comparison with morphological types
from the literature for 450 of our galaxies shows that the reliability
of our classification is quite comparable to that of other
classifiers. We recalibrate the ENACS spectral classification with the
new morphological types, and find that early- and late-type galaxies
can be distinguished from their spectra with 83\%
reliability. Ellipticals and S0 galaxies can hardly be distinguished
on the basis of their spectra, but late spirals can be classified {\em
from the spectrum alone} with more than 70\% reliability. \\ We derive
pseudo-colors and linestrengths from the ENACS spectra for the
galaxies of different morphological types. We consider the bright
($M_R \le -20$) and faint ($M_R > -20$) subsets of the galaxies
without emission lines (non-ELG) separately. We find a strong and
significant correlation between the average color and the average
strength of the metal absorption lines. The average metallicity
decreases and the average color gets bluer towards later Hubble
type. Also, the faint galaxies in {\em each} morphological class are
bluer and less metal-rich than their brighter counterparts, which
extends the well-established color-magnitude relation of early-type
galaxies to (late) spirals. In view of these very strong global
trends, the colors and metallicities of faint S0 galaxies and bright
early spirals are remarkably similar. The bright early spirals may, on
average, have somewhat stronger H$\delta$ absorption than the other
galaxies, which could be due to recent starformation. \\ The galaxies
with emission lines (ELG) have a bluer spectral continuum than the
non-ELG, and the amount of blueing hardly depends on morphological
type. The fraction of ELG depends strongly on morphological type
(varying from $4\pm1$\% for ellipticals to $59\pm4$\% for late
spirals), but for each of the morphological types it varies very
little with projected distance from the cluster center.
}

\maketitle

\section{Introduction}
\label{s-intro}

After the introduction by Hubble of his galaxy classification scheme
(1926, 1936), various other schemes have been proposed, e.g. by Morgan
(1958), de Vaucouleurs (1959), Sandage (1961) and van den Bergh
(1960a, 1960b, 1976). Although the details of those schemes differ,
they all share Hubble's original idea that in some way the
morphological classification reflects an underlying physical
sequence. In individual cases it may be difficult to estimate the
morphological type accurately from visual inspection, but the results
of expert classifiers agree quite well (Naim et al.\ 1995a). Although
it is generally accepted that the morphological types form a sequence,
it is not totally clear how sharp the boundaries are between
successive types, nor whether the galaxy types indicate fundamental
differences in several physical properties, or just a continuous
change in a single physical quantity, such as e.g. bulge fraction (see
van den Bergh 1997).

Studies of the color-magnitude relation (e.g. Bower et al.\ 1992;
Ellis et al. 1997) and of mass-to-light ratios (e.g.\ Lucey et al.\
1991; van Dokkum et al.\ 1996, 1998; Kelson et al.\ 2000) suggest that
ellipticals and S0 galaxies form a single class of slowly evolving
objects. From spectra, Jones et al.\ (2000) conclude that ellipticals
and S0 galaxies have similar luminosity-weighted ages and metal
abundances. However, Kuntschner \& Davies (1998) obtained shorter
luminosity-weighted ages for S0 galaxies in the Fornax cluster, and
Smail et al. (2001) found shorter luminosity-weighted ages for the
{\em fainter} S0 galaxies. At the same time, Dressler et al.\ (1997)
and Fasano et al.\ (2000) found evidence that S0 galaxies have formed
relatively recently (but see Andreon 1998 and Fabricant et al.\ 2000),
and Poggianti et al.\ (1999), and Jones et al.\ (2000) even suggested
that early spirals have transformed into S0 galaxies.

Poggianti et al.\ (1999) used the spectral catalogue of Dressler et
al.\ (1999), of galaxies at intermediate redshifts ($z\sim0.5$), to
study the connection between morphological and spectral
transformations. They distinguish the following (spectral) galaxy
types: starforming emission-line galaxies (ELG), post-starburst
galaxies (a+k/k+a) and passively evolving galaxies (k). The fraction
of post-starburst galaxies increases with redshift (Dressler \& Gunn
1983; Couch \& Sharples 1987). Poggianti et al.\ (1999) suggested that
when individual galaxies entered clusters at intermediate redshifts
they had a modest final starburst (see also Ellis et al.\ 2001;
Ellingson et al.\ 2001) after which the galaxy evolved passively.
These results agree with the finding that the fraction of gas-rich
galaxies and the fraction of ELG decreases towards the cluster center
(Biviano et al.\ 1997, paper III; Solanes et al. 2001; Dale et al.\
2001).

The transformation of a galaxy with starformation or even a modest
starburst to a galaxy that is passively evolving will lead to changes
in the structure of the galaxy. However, Poggianti et al. (1999)
concluded that the morphological transformation, from early spiral to
S0, cannot be driven purely by the star-formation history.  Yet, the
transformation of a significant number of early spirals into S0
galaxies, may be accompanied by (possibly subtle) spectral differences
due to recent star-formation. Therefore, it is of interest to study
the average spectra of galaxies of different morphological types,
using large samples of cluster galaxies.

We have determined the morphological types of close to 2300 galaxies
in the ESO Nearby Abell Cluster Survey (ENACS, Katgert et al.\ 1996,
1998, papers I and V of this series) from CCD images taken especially
for this purpose. The majority of those galaxies (about 1450) had
earlier been classified on the basis of their ENACS spectrum by de
Theije \& Katgert 1999 (paper VI). Following previous work (e.g.\
Zaritsky et al.\ 1995; Folkes et al.\ 1996; Lahav et al. 1996) we
combined the morphological and spectral classifications into a single
classification scheme. This is shown to be meaningful, and it
increases the number of galaxies with `type'-information, which is
essential for the study of morphological segregation (Thomas
\& Katgert 2005, paper IX; Biviano et al. 2002, paper XI). Finally, 
we study the average spectral properties of the galaxies in the
various morphological classes.

This paper is organized as follows. In \S~\ref{s-data} we summarize
the data. In \S~\ref{s-morphology} we investigate the consistency
between our morphological types and those from the literature. In
\S~\ref{s-elges} we investigate the relation between galaxy morphology
and the presence of emission lines in the spectrum. In
\S~\ref{s-sp_class} we compare the spectral and morphological types,
and in \S~\ref{s-combin} we discuss the combination of spectral and
morphological types into one classification scheme. In
\S~\ref{s-spectra} we compare the average spectra for the galaxies of
the various morphological classes. Finally, we summarize and briefly
discuss the main results in \S~\ref{s-discuss}.

\section{The data}
\label{s-data}

Our analysis is based on data from the ENACS, which has yielded
redshifts for 5634 galaxies in the directions of 107 nearby ($z \la
0.1$) Abell cluster candidates (see papers I and V). The ENACS
magnitudes are R-band, and the absolute magnitudes, $M_R$, were
derived for $H_0 = 100$ kms$^{-1}$ Mpc$^{-1}$. Interlopers
(non-members) were eliminated with the interloper removal procedure
devised by den Hartog \& Katgert (1996) and applied to the ENACS
dataset by Katgert et al. (2004, paper XII). For 3798 galaxies in the
ENACS sample, de Theije \& Katgert (paper VI) used the ENACS optical
spectrum to derive spectral galaxy types. The latter were `calibrated'
with a subset of the galaxies for which Dressler (1980) had previously
derived a morphological type. The spectral classification is based on
the wavelength range from 372 nm to 501.4 nm, which was observed at
$\sim0.6$ nm resolution and sampled at 0.35 nm. For a detailed
description of the PCA/ANN-based spectral classification we refer to
paper VI.

In recent years, a long-term programme of CCD-imaging has been carried
out with the 0.92-m Dutch telescope at La Silla. This has yielded
R-band CCD-images for 2295 ENACS galaxies, from which we have derived
morphological types. For $\sim1450$ of those 2295 galaxies we can now
compare the spectral classification from paper VI with the new,
CCD-based morphological type (this work). We must stress that the
galaxies for which we have CCD-images are not a representative subset
of the ENACS sample, because we selected relatively more early-type
galaxies for our CCD-imaging programme.

The CCD that was used in the imaging had pixels of
$0.44^{\prime\prime}$. Most observations were done in photometric
conditions, and the seeing was between $1.2^{\prime\prime}$ and
$2.0^{\prime\prime}$ (FWHM). To get reasonable signal-to-noise ratios
for the galaxies, of which the apparent magnitudes in R are brighter
than 17, exposures of 3 minutes were made. Galaxies in the clusters
A2361, A2401 and A2569 were observed with the 1.0-m JKT at La
Palma. The latter observations were done with a similar resolution and
seeing, but in non-photometric conditions.

\section{The morphological classification and its reliability}
\label{s-morphology}

\begin{figure*}
\includegraphics[width=18.0cm]{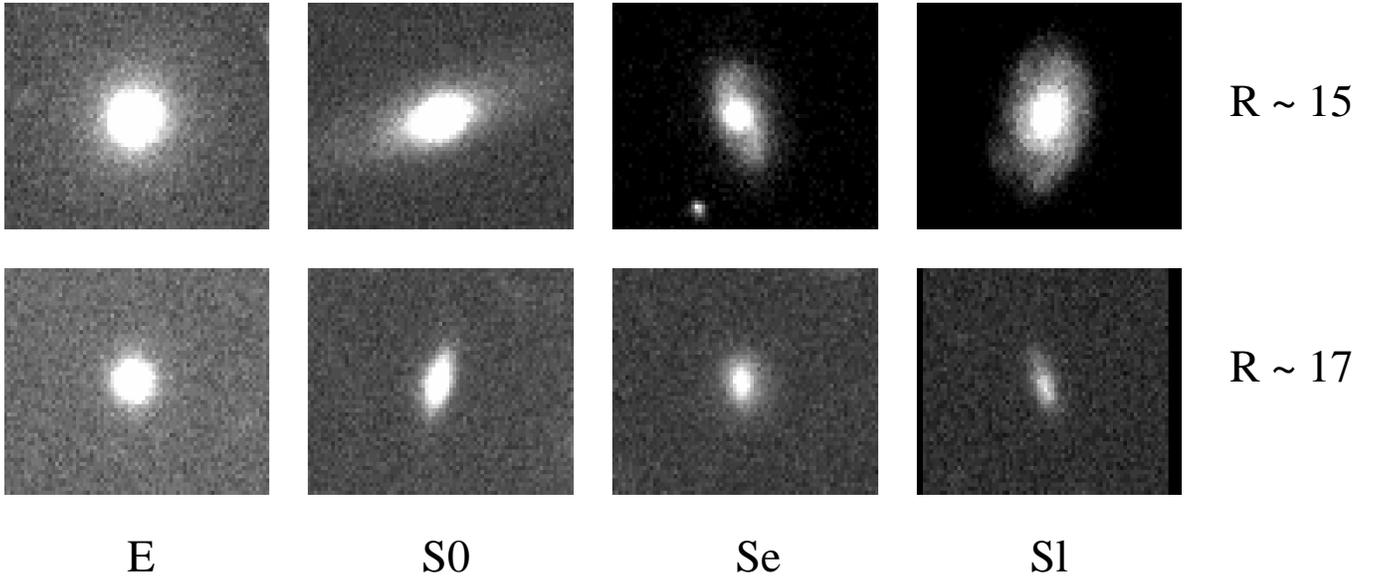}
\caption[]{Images of representative galaxies of all 4 morphological 
types, at two apparent magnitudes. Note that the grey scale contrasts
are not identical in the 8 frames. As we explained in the text, for
the fainter galaxies we sometimes used the brightness profile as
additional evidence for the morphological classification.}
\label{f-images}
\end{figure*}

We used the revised Hubble scheme (de Vaucouleurs 1959, 1963) to
classify the galaxies by eye from their morphological appearance. The
angular resolution of our data allowed us to assign galaxies to one of
the following classes: E, E/S0, S0, S0/S, and S. The E/S0 and S0/S
types are intermediate between E and S0, and S and S0
respectively. The spirals (S) are subdivided into early spirals (Se),
late spirals (Sl) and generic spirals (Sg). In our classification
scheme the Se class comprises Sa, Sab and Sb galaxies, and the Sl
class Sbc, Sc, S/I and I (Irregular) galaxies. The Sg class consists
of galaxies which we know are spirals, but for which no detailed
classification was possible.

In our classification we used the criteria laid down by Dressler
(1980). E galaxies have smooth (de Vaucouleurs) radial profiles
without discontinuities in surface brightness. S0 galaxies have an
additional non-spheroidal component (disk or lens). Spiral galaxies
are (exponential) disk galaxies with spiral arms or ring patterns. The
early spirals have brighter bulges and less open spiral arms than late
spirals. The class of late spirals also contains irregular galaxies,
i.e. amorphous objects with asymmetric isophotes.

Although the criteria used in the Hubble scheme are widely accepted,
different authors are likely to have weighted these criteria in
slightly different ways. In addition, the morphological classification
is not very quantitative, and it can depend on the inclination and the
brightness of the galaxy. In particular, face-on S0 galaxies are
difficult to distinguish from E galaxies. It is also difficult to
distinguish edge-on S0 galaxies and early spirals since our images in
general do not allow us to make statements about the presence of gas
or dust. Likewise, early and late spirals are hard to distinguish if
seen edge-on. In some cases, we used the 1-d brightness profile to
help discrimate between E and S0 classification on the one hand and S0
and spiral classification on the other hand.  In Fig.~\ref{f-images}
we show images of representative galaxies of all 4 morphological
types, at two apparent magnitudes. Note that the grey-scale contrasts
are not identical in the 8 frames.

Naim et al.\ (1995a) made a comparative study of independent
morphological classifications by various expert classifiers, in order
to investigate the performance of their own, automated classification
(Naim et al.\ 1995b). They numbered the successive broad Hubble
classes E, S0, Sa+Sb, Sc+Sd and Sm+I, from 1 to 5 respectively, and
determined the (cumulative) rate of agreement between different
classifiers to within $n$ broad Hubble types. Among others, they
compared Dressler's (1980) classification with both Buta's and
Huchra's classifications. They concluded that the rates of agreement
between classifications of different observers, and between those
classifications and their automated classification, are very similar.

\begin{table}[!b]
\centering
\caption[]{Number of galaxies per Hubble class; AD vs. TK}
\begin{tabular}{|cc|rrrrrrr|}
\hline
      &      & \multicolumn{7}{c|}{TK} \\
      &      & E & E/S0 & S0 & S0/S & Sg & Se & Sl \\
\hline
      &  E            &  71 & 12 & 11 &  1 &  0  & 0 & 0 \\
      & E/S0          &  13 &  4 &  6 &  0 &  0  & 0 & 0 \\
      &  S0           &  18 & 19 &106 & 25 &  6  & 2 & 1 \\
AD    & S0/S          &   0 &  0 & 12 &  7 &  0  & 4 & 0 \\
      &  Sg           &   0 &  0 &  2 &  9 & 16  & 19 & 6 \\
      &  Se           &   0 &  0 &  1 &  9 &  3  & 47 & 2 \\
      &  Sl           &   0 &  0 &  0 &  0 &  1  &  3 & 12 \\
\hline
\end{tabular}
\label{t-AD_TK}
\end{table}

We compared our classification (TK) with that of Dressler (AD), for
the 448 galaxies classified both by Dressler and by ourselves. In
addition to these 448 galaxies, there are 138 galaxies with literature
morphologies and with ENACS redshifts, but for which we do not have a
CCD image. The numbers of galaxies in the AD-TK comparison are listed
in Tab.~\ref{t-AD_TK}. According to Tab.~\ref{t-AD_TK} the total rate
of exact agreement between AD and TK (i.e., using all classes) is
$59\pm 4$\%. Excluding the E/S0- and S0/S-classes because these are
not very meaningful for this comparison, we obtain $77\pm 5$\%.
Finally, if we also exclude the indeterminate Sg-class comparisons
(i.e. Sg-Sg, Sg-Se and Sg-Sl), we obtain $79\pm 5$\% agreement. This
latter number is the relevant one for all analyses, in which the S/S0,
E/S0 and Sg classes are not used.

\begin{figure}[!t]
\includegraphics[width=8.8cmcm]{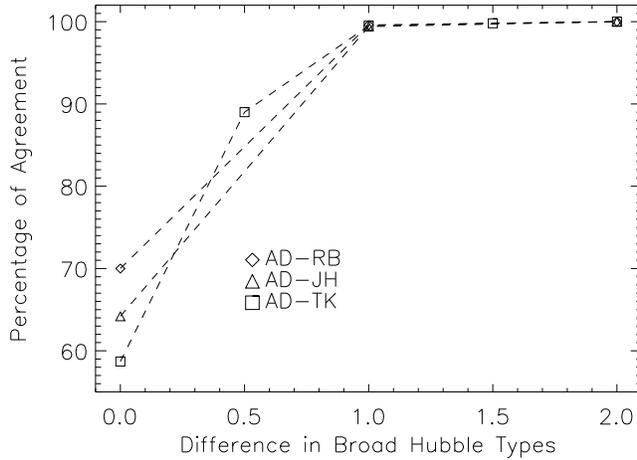}
\caption[]{The cumulative rate of agreement between the results
of different classifiers, to within $n$ broad Hubble types. Clearly,
the performance of the different classifications is very similar.}
\label{f-clagree}
\end{figure}

In Fig.~\ref{f-clagree} we show the (cumulative) rate of agreement to
within $n$ broad Hubble types, where we followed Naim et al. (1995a)
and numbered the successive broad Hubble types as follows: E=1.0,
E/S0=1.5, S0=2.0, S0/S=2.5, Sa+Sab+Sb=3.0, Sg=3.5, Sbc+Sc=4.0 and
S/I+I=5.0. We show the rates of agreement between Dressler and Buta
(AD-RB), between Dressler and Huchra (AD-JH), and between AD and TK,
as a function of the difference between the two types. The 4
classification systems are seen to have very similar performance, with
more than 80\% of the comparisons showing a difference of not more
than half a broad Hubble type.

For those 185 galaxies in common between AD and TK for which the two
types are not identical (see Tab.~\ref{t-AD_TK}), we adopted our
morphological type, except in the following cases: (1) if the galaxy
was classified as E/S0 by us and as E or S0 by AD, (2) if the galaxy
was classified as S0/S by us and as S0 or Se by AD, or (3) if the
galaxy was classified as Sg by us and as Se or Sl by AD. In addition,
we adopted the morphological type given by Dressler (or by others) for
those 138 galaxies in the ENACS for which we do not have a CCD image
and thus, no morphological type.

\section{The emission-line galaxies (ELG)} 
\label{s-elges}

Biviano et al.\ (1997, paper III) discussed the presence of the [OII]
(372.7 nm), $H\beta$ (486.1 nm) and the [OIII] doublet ($495.9+500.7$
nm) emission lines in the ENACS spectra. In the ENACS galaxies, these
emission lines in general indicate the presence of warm gas, which is
most likely connected to starformation. Galaxies with one or more of
these emission lines are referred to as ELG.

In Tab.~\ref{t-ELG_morph} we show the distribution of the ELG over
morphological type for all galaxies in the ENACS sample. The fraction
of ELG is very low for early-type galaxies ($4\pm1$\% for E galaxies
and $5\pm1$\% for S0 galaxies) and rises from $27\pm4$\% for the early
spirals to $59\pm4$\% for the late spirals. The latter implies that
ELG and late spirals are very closely related. This is also evident in
the analysis of the projected phase-space distributions, as described
in paper XI. ELG and late spirals without emission lines have very
similar distributions, while they are significantly different from
those of other galaxy types.

\begin{figure}
\includegraphics[width=8.8cm]{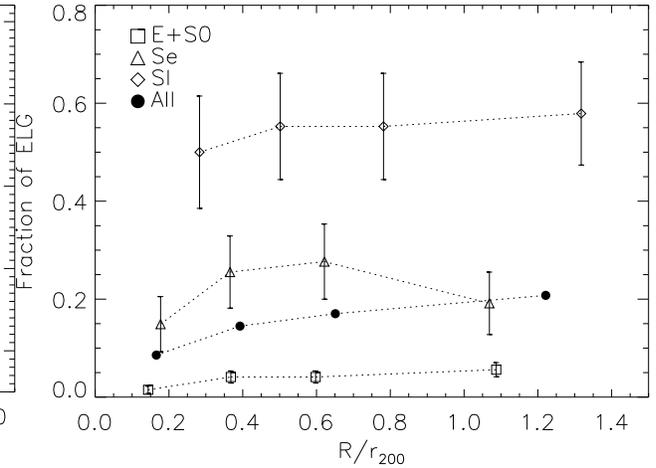}
\caption[]{The dependence of the fraction of ELG on the normalised 
projected cluster-centric distance $R/r_{200}$, for all galaxies in
the 67 clusters with at least 20 member galaxies and with $z<0.1$
(filled circles). Also shown are the fractions for the individual
galaxy classes.}
\label{f-elg_rad}
\end{figure}

Because galaxies of different types have different projected
distributions, one would expect the ELG and non-ELG to have different
radial distributions. In Fig.~\ref{f-elg_rad} we show the relation
between the fraction of ELG and the projected distance from the
cluster center, for the early-type (E+S0) galaxies, the early spirals,
the late spirals and all types together. We used the galaxies in the
67 clusters with at least 20 member galaxies and with $z<0.1$.
Projected distances $R$ were scaled with $r_{200}$, the radius within
which the average density of the cluster is 200 times as large as the
critical density of the Universe, and which is a good approximation of
the virial radius (e.g. Carlberg et al. 1997).

\begin{table}[!b]
\centering
\caption[]{The distribution of ELG and non-ELG over morphological type.}
\begin{tabular}{|l|r|r|r|}
\hline
Morphological & ELG & non-ELG & ELG-fraction \\ type & & & \\
\hline
E    &  16 & 344 & 0.04 \\
E/S0 &   9 & 139 & 0.06 \\
S0   &  40 & 763 & 0.05 \\
S0/S &  45 & 241 & 0.16 \\
Sa   &  26 &  92 & 0.22 \\
Sab  &  10 &  25 & 0.29 \\
Sb   &  26 &  54 & 0.32 \\
Sg   &  92 & 149 & 0.38 \\
Sbc  &  12 &  13 & 0.48 \\
Sc   &  33 &  42 & 0.44 \\
S/I  &  34 &  22 & 0.61 \\
I    &  54 &  14 & 0.79 \\
\hline 
All  & 397 & 1898 & 0.17 \\
\hline
\end{tabular}
\label{t-ELG_morph}
\end{table}

We used the Kolmogorov-Smirnov (or KS-) test to estimate the
probability that the radial distributions of ELG and non-ELG are drawn
from the same parent distribution. According to this test the radial
distributions of ELG and non-ELG are different ($P_{KS}<0.01$); i.e.
the fraction of ELG decreases significantly towards the center.
However, if we consider the Se or the Sl galaxies separately, the
fraction of ELG does not vary with $R$, so there is no significant
difference between the distributions of ELG and non-ELG for these
classes. Yet, there is a significant difference between the overall
fractions of ELG for Se and Sl galaxies. This is in agreement with the
result of Solanes et al. (2001), who concluded that gas is expelled
more easily from early spirals than from late spirals. Apparently, the
efficiency of the gas removal is more related to the type of
spiral than to its position in the cluster. 

\section{The spectral classification revisited}
\label{s-sp_class}

De Theije \& Katgert (paper VI) used a principal-component analysis
(PCA) in combination with an artificial neural network (ANN) to
classify most of the ENACS galaxies on the basis of their ENACS
spectrum. They used the first 15 principal components in the ANN, of
which the first describes the slope of the spectrum, and the second
one the curvature; the following principal components describe several
other features of the spectrum. The ANN was applied to a training set
of 150 of the 270 galaxies for which morphological types (E, S0 or
S+I) were determined by Dressler (1980). To each combination of the 15
principal components, a spectral type (E$^*$, S0$^*$ or (S+I)$^*$)
could be assigned, in such a way that the spectral and morphological
types matched as closely as possible.

De Theije \& Katgert estimated the reliability of their spectral
classification by repeating the ANN routine for 10 different, but --
necessarily -- correlated training sets. They thus obtained, for each
galaxy, three probabilities; viz. $P_{\rm E^*}$, the number of times
out of 10 that the galaxy was classified as an elliptical (E$^*$),
$P_{\rm S^*}$, the number of times out of 10 that the galaxy was
classified as (S+I)$^*$ and $P_{\rm S0^*}$ ($= 1 - P_{\rm E}^* -
P_{\rm S}^*$) the number of times out of 10 that the galaxy was
classified as an S0$^*$.

\begin{figure}
\includegraphics[width=8.8cm]{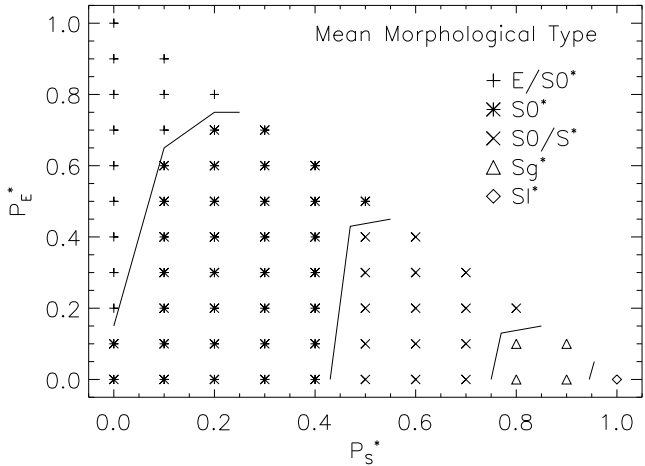}
\caption[]{For each of the 66 combinations of $P_{\rm E^*}$ and 
$P_{\rm S^*}$, as determined by de Theije \& Katgert from a PCA/ANN
analysis of the ENACS spectra, we show the `mean' morphological type
(see text).}
\label{f-spclass}
\end{figure}

Because the CCD-imaging program has greatly increased the number of
galaxies with morphological types, we could check and recalibrate the
PCA/ANN classification scheme of paper VI with much better statistics.
For each combination of $P_{\rm {E^*}}$ and $P_{\rm {S^*}}$ we derived
the `mean' morphological type, which we will later use as the
`spectral' type for that ($P_{\rm E^*}$, $P_{\rm S^*}$)-combination.
However, in doing so we forced transitions from an earlier to a later
type to occur only when $P_{\rm S^*}$ increases or $P_{\rm E^*}$
decreases. In Fig.~\ref{f-spclass} we show how the types, E/S0$^*$,
S0$^*$, S0/S$^*$, Sg$^*$, and Sl$^*$, vary with $P_{\rm E^*}$ and
$P_{\rm S^*}$. It is clear from this Figure that $P_{\rm S^*}$ is the
main factor driving the classification, while $P_{\rm E^*}$ appears to
be only of secondary importance.

In Tab.~\ref{t-morph_spectr} we show the relation between the
morphological and spectral types, for the 1071 galaxies for which both
types are available. Note that the Table does not include galaxies
with morphological types E/S0 or S0/S (about 220), while galaxies with
a spectral type S0/S$^*$ (about 160) were not included either.

\begin{figure}
\includegraphics[width=8.8cm]{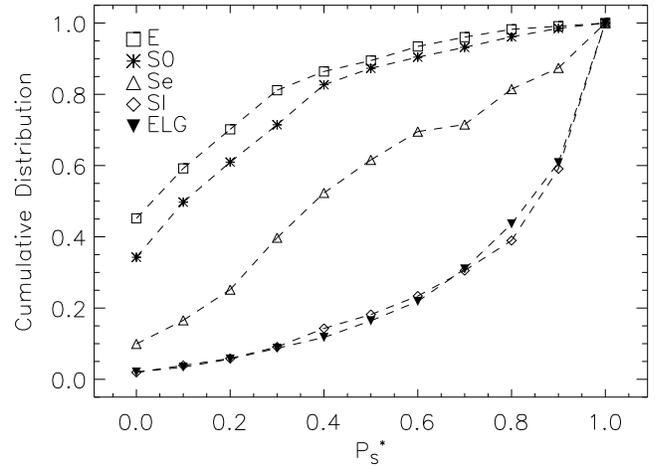}
\caption[]{The normalized cumulative $P_{\rm S}^*$-distributions 
of the 4 morphological galaxy classes and of the emission-line
galaxies.}
\label{f-psdistr}
\end{figure}

For 647 galaxies, the morphological and spectral types agree that the
galaxy is of early type (i.e. E, E/S0 or S0), while for 240
galaxies both types indicate the galaxy to be of late type. So, the
spectral classification can distinguish early-type and late-type
galaxies with $83\%$ reliability. 

\begin{table}[b]
\centering
\caption[]{The relation between morphological and spectral galaxy 
           types.}
\begin{tabular}{|cc|rrrrr|}
\hline
  & & \multicolumn{5}{|c|}{Morphological Type} \\ & & E & S0 & Se & Sg
         & Sl \\
\hline
         & E/S0$^*$  & 101 & 170 & 15  &  3  &  3  \\
Spectral &  S0$^*$   &  96 & 280 & 64  & 35  & 19  \\
  Type   &  Sg$^*$  &   7 &  28 & 23  & 47  & 44  \\
         &  Sl$^*$   &   2 &   8 & 19  & 44  & 63  \\
\hline
\end{tabular}
\label{t-morph_spectr}
\end{table}

From Fig.~\ref{f-spclass} it is clear that the spectral classification
has much more discriminating power for late-type galaxies than for
early galaxy types. As a matter of fact, the `earliest' spectral type
is not E$^*$ but E/S0$^*$. In the E/S0$^*$ class, the morphological
content is actually dominated by S0's (58\%) rather than by E's
(35\%), but the fraction of E's in the E/S0$^*$ class is considerably
larger than in the S0$^*$ spectral class (for which it is 19\%). The
contamination by morphologically classified late-type galaxies in the
E/S0$^*$ class is very small, viz. only $7\pm2$\%. The S0$^*$ class is
purer than the E/S0$^*$ class, although contamination by ellipticals
(19\%) and late-type galaxies (24\%) is far from negligible.

The late spectral types are divided in the Sg$^*$ and Sl$^*$ spectral
classes, because our spectral classification scheme is not able to
identify early spirals with any confidence. The Sg$^*$ class contains
more late spirals than early spirals, but the contamination of
early-type galaxies is also non-negligible, at $23\pm2$\%. The purest
of all spectral classes is the Sl$^*$ class. If we distribute the
generic spirals over the early and late spiral classes, in the
observed proportion of the morphological early and late spirals, we
find that $72\pm5$\% of the galaxies in the Sl$^*$ class have Sl
morphologies.

\begin{table*}
\centering
\caption[]{The catalogue of galaxy types (extract; the full catalogue 
           is available electronically).}
\begin{tabular}{|cc|ccc|c|cc|ccc|c|}
\hline
Cluster & galaxy & \multicolumn{3}{|c|}{galaxy type} & ELG & Cluster &
galaxy & \multicolumn{3}{|c|}{galaxy type} & ELG \\ name & nr.  &
morph. & spectral & adopted & & name & nr.  & morph. & spectral &
adopted & \\
\hline
A0013 &  1 &  S0/S & S0   & S0   &   & A0013 & 23 &      & Sl   & Sl   & * \\
A0013 &  2 &       & S0   & S0   &   & A0013 & 24 & S0   & E/S0 & S0   &   \\
A0013 &  3 &       & Sg   & Sg   &   & A0013 & 25 & S0   & S0   & S0   &   \\
A0013 &  4 &  S0   & S0   & S0   &   & A0013 & 26 &      & S0/S & S0/S & * \\
A0013 &  5 &       & Sl   & Sl   & * & A0013 & 27 & S0   & E/S0 & S0   &   \\
A0013 &  6 &  S0   & S0   & S0   &   & A0013 & 28 &      & Sg   & Sg   &   \\
A0013 &  7 &       & S0   & S0   &   & A0013 & 29 & S0   & E/S0 & S0   &   \\
A0013 &  8 &  E    & E/S0 & E    &   & A0013 & 30 & S0   & E/S0 & S0   &   \\
A0013 &  9 &       & S0/S & S0/S &   & A0013 & 31 &      & Sl   & Sl   &   \\
A0013 & 10 &       & Sl   & Sl   & * & A0013 & 32 & S0   & S0   & S0   &   \\
A0013 & 11 &  Sa   & S0   & Se   &   & A0013 & 33 & S0   & S0   & S0   &   \\
A0013 & 12 &  E    & E/S0 & E    &   & A0013 & 34 & Sa   & Sl   & Se   & * \\
A0013 & 13 &  S0   & E/S0 & S0   &   & A0013 & 35 & E/S0 & E/S0 & E/S0 &   \\
A0013 & 14 &       & S0   & S0   &   & A0013 & 36 & S0   & S0   & S0   &   \\
A0013 & 15 &  Sa   & S0   & Se   &   & A0013 & 37 &      & Sg   & Sg   &   \\
A0013 & 16 &  S0   & S0   & S0   &   & A0013 & 38 & E    & S0   & E    &   \\
A0013 & 17 &  S0   & S0   & S0   &   & A0013 & 39 & S0/S & E/S0 & S0   &   \\
A0013 & 18 &       & S0/S & S0/S &   & A0013 & 40 &      & Sg   & Sg   &   \\
A0013 & 19 &  E    & E/S0 & E    &   & A0013 & 41 & Sg   & Sl   & Sl   & * \\
A0013 & 20 &  E/S0 & E/S0 & E/S0 &   & A0013 & 42 &      & Sl   & Sl   & * \\
A0013 & 21 &       & S0/S & S0/S &   & A0013 & 43 &      & S0/S & S0/S & * \\
A0013 & 22 &       & S0   & S0   &   & A0013 & 44 &      & S0   & S0   &   \\
\hline
\end{tabular}
\label{t-classif}
\end{table*}

Because the fraction of ELG varies with morphological type, we also
compared the $P_{\rm S^*}$-distributions of all morphological classes,
separately for non-ELG and ELG. The $P_{\rm S^*}$-distributions of all
4 non-ELG morphological classes (E, S0, Se and Sl) are different
according to KS tests (i.e. have less than 1\% probability to be drawn
from the same distribution). The $P_{\rm S^*}$-distributions of the 3
ELG morphological classes (we did not consider the 13 ELG ellipticals)
are not all different. The early and late spirals with emission lines
do not have significantly different $P_{\rm S^*}$-distributions, but
this negative result is probably due to limited statistics.

Although it is impossible to make a reliable distinction between
ellipticals and S0 galaxies on the basis of the spectrum alone,
ellipticals have somewhat earlier spectral types than S0 galaxies.
This is evident in Fig.~\ref{f-psdistr}, which shows the $P_{\rm
S^*}$-distributions of the various morphological classes.  Although
the $P_{\rm S^*}$-distributions of the E and S0 galaxies are very
similar, a KS-test shows them to have less than $1$\% probability to
be drawn from the same parent distribution. From Fig.~\ref{f-psdistr}
it is also clear that the late spirals and the ELG are
spectroscopically indistinguishable, while the spectra of both are
very different from those of early spirals and E and S0 galaxies. The
great spectral similarity of late spirals and ELG has led us to
classify all generic spectroscopic spirals (i.e. Sg$^*$) with emission
lines as Sl$^*$. By the same token, ELG for which no PCA/ANN spectral
type could be obtained were classified as Sl$^*$.

\section{Combining morphological and spectral types}
\label{s-combin}

For 2433 galaxies we have a morphological type: 1847 of those had the
type determined only from the present CCD imaging; another 138 only
have a type from the literature, while for the remaining 448 there are
morphological types from the literature as well as from the present
CCD imaging. For 185 of the latter, the two morphological types were
not identical (although often consistent), and for those we adopted
our own type, except in the few cases described at the end of
\S~\ref{s-morphology}.

Revised spectral types (see \S~\ref{s-sp_class}) were available for
3798 galaxies. In addition there are 421 galaxies for which no
spectral type was available (primarily because the de-redshifted
wavelength interval of the ENACS spectrum did not have sufficient
overlap with the standard range adopted in the PCA/ANN analysis -- see
paper VI), but for which the spectrum was inspected and found to
contain one or more emission lines (see paper III). This latter (ELG)
information is less specific about galaxy type than the PCA/ANN
spectral type. Yet, it was sometimes used to assign the spectral type
Sl$^*$, viz. if there was no PCA/ANN spectral type, or if the spectral
type was Sg$^*$.

We have combined the morphological and (recalibrated) spectral types
into a single set of 4879 galaxy types in the following way. For about
half of the galaxies, we only have a spectral type and that was then
adopted as the galaxy type. For the other half, we had a morphological
type and sometimes a spectral type. For those, the morphological type
was adopted, except in those cases where the spectral type could help
resolve an ambiguous or non-specific morphological type, like E/S0,
S0/S or Sg. 

If the morphological type was E/S0 and the spectral type was S0$^*$
the galaxy was classified as S0. If the morphological type was S0/S
and the spectral type was E/S0$^*$ or S0$^*$ the galaxy was also
classified as S0. If the morphological type was S0/S and the spectral
type was Sg$^*$, while the galaxy had $P_{\rm S^*} < 0.9$, it was
classified as Se. Similarly, we tried to make the morphological Sg
type more specific, as follows. If the spectral type was E/S0$^*$,
S0$^*$ or S0/S$^*$, the adopted galaxy type was Se because 75\% of the
spirals in these spectral classes have Se morphologies. Likewise, if
the spectral type was Sl$^*$ or Sg$^*$ with $P_{\rm S}^* \ge 0.9$, the
adopted galaxy type was Sl, because 77\% of the spirals in this
spectral class have Sl morphologies.

This procedure yields a consistent classification for 4879 ENACS
galaxies. For 2102 (i.e 43\%) of those, the morphological type was
adopted; for 2446 (i.e 50\%) the spectral type was adopted, while for
the remaining 331 (i.e. 7\%) of the galaxies, the morphological and
spectral types were combined, as described above.

The results for the 4879 galaxies are available at the CDS in
Strasbourg (http://cdsweb.u-strasbg.fr/Abstract.html). The final
galaxy types that we used are: E, E/S0 (could be either), S0, S0/S
(could be either), Se (could be Sa, Sab, Sb, and sometimes the
morphological type is indeed more specific), Sl (could be Sbc, Sc, S/I
and I, and sometimes the morphological type is indeed more specific)
and Sg (generic spiral without information on subtype). In
Table~\ref{t-classif} we show, as an example, the results for the 44
galaxies in cluster Abell 13. For each galaxy we give the number (as
in the original ENACS catalogue -- see Katgert et al. 1998, paper V),
the morphological and spectral galaxy type, and the ELG character
(when applicable) together with the final galaxy type adopted in
subsequent papers, obtained as described above.

\section{The average spectral properties of the various galaxy classes}
\label{s-spectra}
            
The PCA appears to be a good tool to distinguish between spectra, but
the physical meaning of most of the principal components, and
especially of the higher-order ones, is not intuitively clear. We have
tried to get a better understanding of the spectral differences
between the various morphological types by studying the average
spectra of the different galaxy classes.
      
\begin{figure}
\includegraphics[width=8.8cm]{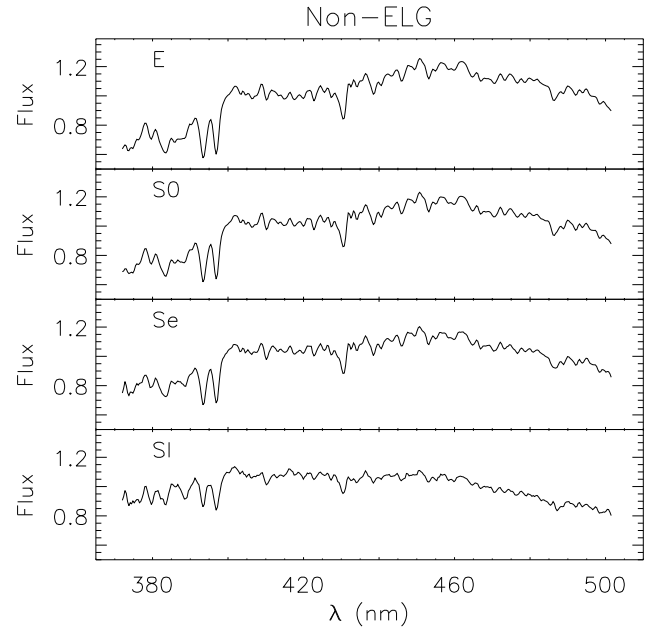}
\caption[]{The mean spectra for ellipticals (E), S0 galaxies, 
early spirals (Se) and late spirals (Sl), without emission lines
(non-ELG). The mean spectrum was determined by averaging individual
spectra in which the mean flux was normalised to one (arbitrary
units). The individual galaxy spectra were {\it not}
spectro-photometrically calibrated, but have been corrected for the
spectral response of spectrograph and detector.}
\label{f-avspnelg}
\end{figure}

\subsection{The average spectra of the non-ELG}
\label{s-spectra_av}
      
In Fig.~\ref{f-avspnelg} we show the mean spectra for the 4 non-ELG
classes. The mean spectrum for each class was determined by averaging
individual spectra in which the mean flux was normalized to one
(arbitrary units). The spectra were not calibrated
spectro-photometrically, but the spectral response of the grating and
the CCD was corrected as described in paper VI. Although one could
argue that the average of the spectra of many galaxies does not
necessarily have a clear physical meaning, Fig.~\ref{f-avspnelg}
illustrates the well-known global spectral differences between late
spirals and early-type galaxies, both as regards spectral slope and
strength of the absorption lines. 

Because color has been shown to depend on luminosity (e.g.\ Bower et
al.\ 1992), while luminosity-weighted ages were found to increase from
faint to bright S0 galaxies (see e.g.\ Smail et al.\ 2001), we also
compared the average spectra of `faint' and `bright' non-ELG galaxies.
For this comparison, we estimated the magnitude limit between `faint'
and `bright' galaxies in the following way. For various limits in
absolute magnitude we compared the $P_{\rm S^*}$-distributions of
non-ELG galaxies fainter and brighter than this limit. For limits
between $M_R=-19.5$ and $M_R=-20.0$ ($H_0 = 100$ kms$^{-1}$
Mpc$^{-1}$) the ${\rm P_S^*}$-distributions of faint and bright S0
galaxies are different according to KS-tests. Because the number of
`faint' galaxies is relatively small, we chose $M_R=-20$ as the
magnitude limit. In Tab.~\ref{t-nonELG} we give the number of `bright'
and `faint' galaxies for each type.

\begin{table}[!b]
\centering
\caption[]{The number of non-ELG galaxies for which we determined 
morphological type, `colors' and `linestrengths'. }
\begin{tabular}{|c|rrrr|}
\hline
   &   \multicolumn{4}{|c|}{Morphological Type} \\
               & E & S0 &  Se & Sl \\
\hline
Bright ($M_R \le -20$) & 203 & 326 &  91  & 41 \\
Faint  ($M_R > -20$) &  27 & 207 &  25  & 20 \\
\hline
Total  & 230 & 533 & 116 &  61 \\
\hline
\end{tabular}
\label{t-nonELG}
\end{table}

\begin{figure}
\includegraphics[width=8.8cm]{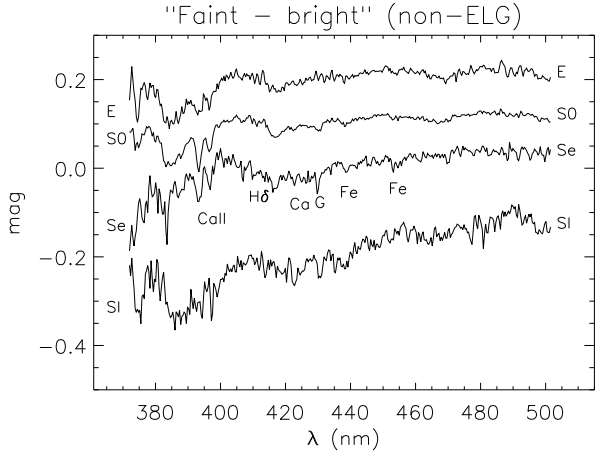}
\caption[]{The difference (in magnitudes) between the spectra of 
the faint ($M_R>-20$) and bright ($M_R \le -20$) non-ELG galaxies, for
different morphologies. The difference spectra have zero mean flux, but
were shifted so that they could be shown in one figure.}
\label{f-ft-brnonelg}
\end{figure}

In Fig.~\ref{f-ft-brnonelg} we show the differences between the mean
spectra of faint ($M_R>-20$) and bright ($M_R \le -20$) galaxies for
the 4 non-ELG classes. A positive slope in this figure corresponds to
a bluer continuum of the first (fainter) galaxy class. Therefore, the
bright galaxies in general have a redder continuum than the faint
ones. This conclusion, which is well-established for the early-type
galaxies (see, e.g., Bower et al.\ 1992), also appears to be true for
the spirals.

We quantified these differences by determining pseudo-colors $''b-v''$
and $''v-g''$, where the $''b''$-, $''v''$- and $''g''$-band
magnitudes were defined as the median magnitudes in the spectral
ranges 372 to 389.5 nm, 405 to 422.5 nm, and 483.9 to 501.4 nm
respectively. It appears that $''b-v''$ and $''v-g''$ are rather
insensitive to the presence of emission lines, but $''b-v''$ is
sensitive to the strength of the 400 nm break. For each galaxy class
we determined the mean colors by averaging the estimates of individual
galaxies. The uncertainty in the mean colors was determined by
dividing the dispersion of the individual estimates within a galaxy
class by the square root of the number of galaxies (see
Tab.~\ref{t-nonELG}). The 'uncertainty' therefore derives primarily
from the intrinsic spread within a class.

\begin{figure}
\includegraphics[width=8.8cm,height=6.5cm]{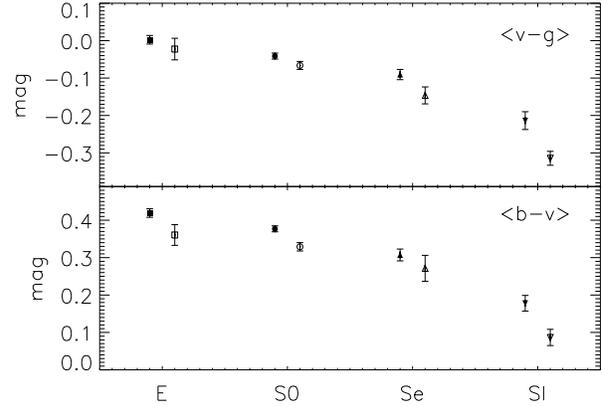}
\caption[]{The mean colors $\langle ''v-g'' \rangle$ (top) and 
$\langle ''b-v'' \rangle$ (bottom) of the various morphological
classes. Note that the colors are instrumental, because the spectra
were not spectro-photometrically calibrated. For each of the 4 galaxy
classes we show the colors for bright ($M_R \le -20$) galaxies (filled
symbols) and faint ($M_R>-20$) galaxies (open symbols) separately.}
\label{f-colcol_nonelg}
\end{figure}

The results are shown in Fig.~\ref{f-colcol_nonelg}, which shows the
mean colors $\langle ''b-v'' \rangle$ and $\langle ''v-g'' \rangle$
for the bright and faint subsets of each of the 4 galaxy classes. The
most obvious effect in this Figure is that the colours get bluer
towards later Hubble type. For the bright galaxies, the effect is
significant at $3\sigma$ or more for all comparisons between
consecutive Hubble classes, and in both colors. For the faint
galaxies, the same trend is visible, albeit that the statistical
weight (and therefore, the significance) is considerably less. The
great similarity between the variation of $\langle ''b-v''\rangle$ and
$\langle ''v-g ''\rangle$ with galaxy type, indicates that the two
colors are strongly correlated. This confirms that the depth of the
400 nm ~break is strongly correlated with the continuum slope.

At the same time, there is a tendency for the bright galaxies in each
class to be redder than the faint ones. The effect is significant at
$3\sigma$ or more for the S0 galaxies in $\langle ''b-v'' \rangle$,
and for the late spirals in both colors. The latter result may be
somewhat unexpected, and in interpreting it one must remember that the
colors (as derived from the ENACS spectra) are valid for a central
region of 3--6 kpc diameter, and thus do not refer to the entire
galaxy. With some exaggeration, one may therefore conclude that the
color-magnitude relation that is well-established for early-type
galaxies even extends to the bulges and the central parts of the disks
of late-type galaxies. Without further data it is impossible to
correct for the disk contribution.

We have also estimated the strengths of the most prominent absorption
lines, viz.: CaII 393.4+396.8 nm, H$\delta$ 410.2 nm, CaI 422.7 nm,
G-band (430.4 nm), FeI 438.3 nm ~and FeI 453.1 nm, and of the
following emission lines: OII 372.7 nm, H$\beta$ 486.1 nm ~and OIII
495.9+500.7 nm. For each line we determined the total flux, $F_l$, in
the central 5 pixels of the line ($\sim1.8$ nm ~or $\sim3$ FWHM of the
instrumental profile), and we determined the continuum flux, $F_c$, in
6 pixels outside these central 5 pixels (3 pixels on each side). We
defined the linestrength as $\Delta m = -2.5\log{F_l/F_c}$.  Because
the CaI 422.7 nm, FeI 438.3 nm ~and FeI 453.1 nm ~lines are relatively
weak, we improved the signal-to-noise ratio by averaging their
strengths into a single value for each galaxy. Note that our $\Delta
m$-values cannot be compared directly to equivalent widths (EW). This
is because we could not fit the line profile with a model (e.g. a
Gaussian) because of the limited resolution of our data.

\begin{figure}
\includegraphics[width=8.8cm]{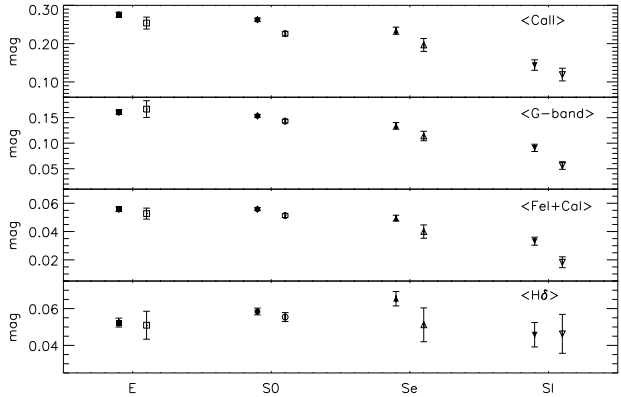}
\caption[]{The strengths of the absorption lines. Filled and open 
symbols refer to the bright and faint subsets of the 4 galaxy classes.
The upper 3 panels show the strengths of the 3 metal line(-complexe)s,
and the lower panel the strength of the H$\delta$ line. For the
definition of the line-strength $\Delta m$, see the text;
absorption-line strength increases upwards.}
\label{f-abslines}
\end{figure}

In the upper 3 panels of Fig.~\ref{f-abslines} we show the average
strengths of the absorption lines (as defined above) for the brighter
and fainter subsets of the different galaxy classes. The CaII, G-band
and FeI+CaI lines are indicators of metallicity. The uncertainty in the
mean linestrengths was determined by dividing the dispersion of the
individual estimates within a galaxy class by the square root of the
number of galaxies (see Tab.~\ref{t-nonELG}). Again, the `uncertainty'
therefore mostly measures the intrinsic spread within a class.

As with the colors, there is a trend in Fig.~\ref{f-abslines} for the
metal-line strengths to decrease towards later Hubble type. However,
the effect is significant at $3\sigma$ or more for the comparison
between early and late spirals only, and in that case it is
significant for all 3 metal lines and for bright as well as faint
subsets. There is also a tendency (certainly among the S0 galaxies and
the early and late spirals) for the metal-lines to be stronger for the
bright galaxies than for the faint ones. The difference is significant
at $3\sigma$ or more for CaII in S0 galaxies only, and for G-band and
FeI+CaI for the late spirals only.

\begin{figure}
\includegraphics[width=8.8cm]{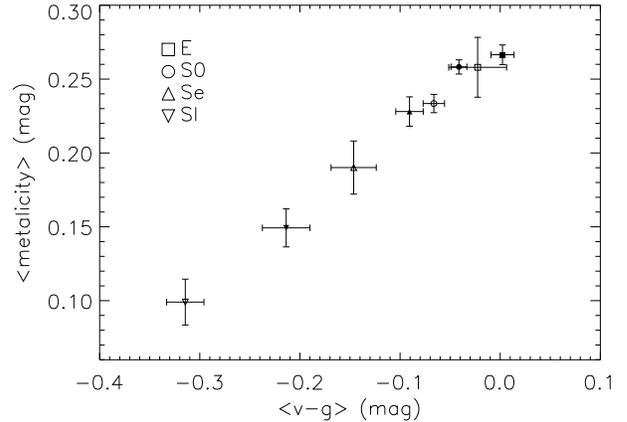}
\caption[]{The relation between color and metallicity. Because the 
$\langle ''v-g'' \rangle$ and $\langle ''b-v'' \rangle$ colors are
very strongly correlated, we only show the correlation with $\langle
''v-g'' \rangle$ color. The metallicity was derived by combining the
three measured line-strengths as described in the text. Note that the
combined metallicity is given on the CaII-scale.}
\label{f-col-abslines}
\end{figure}

In Fig.~\ref{f-col-abslines} we have combined the information on color
and strength of the three metal lines. Because the \mbox{$\langle ''v-g''
\rangle$} and $\langle ''b-v'' \rangle$ colors are very strongly
correlated, we only show the correlation with $\langle ''v-g''
\rangle$ color. For this figure we have chosen to combine the 3 
metal-line strengths. Because the measured strengths of the G-band and
FeI+CaI lines are small, we approximately brought them on the scale of
the CaII linestrength by multiplying them by constant factors
(viz. 1.68 and 4.57 respectively). Subsequently, we averaged the 3
line-strengths to produce the value displayed in
Fig.~\ref{f-col-abslines}.

This Figure summarizes in a compact way the properties of the average
spectra of the non-ELG that we discussed so far. There is a strong and
significant correlation between $\langle ''v-g''\rangle$ color and
metallicity, with a very systematic decrease of metallicity and
blueing of the color towards late Hubble type. In addition, the faint
galaxies in each class are generally bluer and less metal-rich than
the bright galaxies. Although both effects are not always significant
at the $3\sigma$-level in individual comparisons, the general trends
are very strong. One interesting result, which may be significant or
fortuitous, is that colors and metallicities of faint S0 galaxies and
bright early spirals appear to be very similar.

In the bottom panel of Fig.~\ref{f-abslines} we show the strength of
the H$\delta$ absorption line for the different galaxy types, and
luminosity classes. The H$\delta$ absorption line is often used as an
indicator for recent star-formation (e.g.\ Dressler et al.\ 1999).
There are no differences of $3\sigma$ or more between the different
morphological types, neither for the bright nor faint subclasses. Yet,
there may be a tendency for the bright S0 and, in particular, the
bright early spirals to have stronger H$\delta$ absorption than do the
ellipticals and the late spirals. The fact that the bright early
spirals have the strongest H$\delta$ absorption may be due to enhanced
recent starformation that some bright early spirals have undergone
(e.g.\ Poggianti et al.\ 1999).

One may wonder if our result, that the bright early spirals in local
clusters may have a slightly higher H$\delta$ absorption-line strength
is in conflict with the result obtained by Poggianti et al. (2004),
who found that the fraction of luminuous (a+k/k+a) galaxies is much
smaller in the Coma cluster than it is in the $z \sim 0.5$ clusters
studied by Dressler et al. (1999). However, the identification between
their (a+k/k+a) galaxies (spectroscopically defined) and our early
spirals (morphologically defined) is quite doubtful. For one thing,
the average H$\delta$ equivalent width of our (bright) early spirals
is unlikely to be as high as that of the high-redshift (a+k/k+a)
galaxies. The only thing one might conclude is that the fraction of
(a+k/k+a) galaxies is slightly higher among the bright early spirals
than it is on average.

\subsection{The average spectra of the ELG}

\begin{figure}
\includegraphics[width=8.8cm]{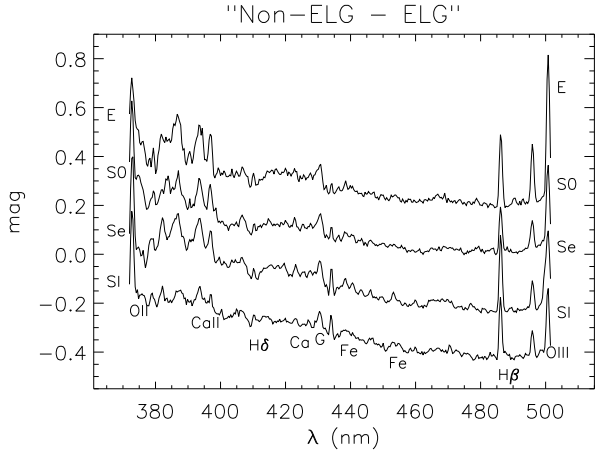}
\caption[fig6]{The ratio (in magnitudes) between the mean non-ELG 
and ELG spectra for different morphological classes. The residual
spectra all have identical (i.e. zero) mean flux, but they were offset
to be shown in one figure.}
\label{f-elg-nonelg}
\end{figure}

In Fig.~\ref{f-elg-nonelg}, we show the differences between the
average spectra of non-ELG and ELG for each of the 4 galaxy classes.
The non-ELG have significantly redder continuum colors and stronger
absorption lines than the ELG ($>3\sigma$-result), and this is true
for all galaxy types. To within the `uncertainties' (which are mostly
due to the spread within a morphological class) the differences
between the mean colors and mean absorption-line strengths of non-ELG
and ELG are indistinguishable for E, S0, Se and Sl galaxies.

This independence of the mean (non-ELG -- ELG) difference spectrum of
morphological type is quite remarkable, even though in second order
there are rather subtle differences in linestrength, while it must
also be kept in mind that the spread within each morphological class
is considerable (with a median spread of $\Delta m \approx 0.08$ for
the non-ELG and $\Delta m \approx 0.10$ for the ELG). Taken at face
value this independence implies that the process that is responsible
for the emission lines appears to produce a blueing of the continuum
of the galaxy, by an average amount that is very similar for all
galaxy types. An explanation for this could be that the primary source
for the excitation of the lines is most likely photo-ionization.

If one combines this result with the conclusions about the
type-dependence of the colors of non-ELG (as summarized in
Fig.~\ref{f-colcol_nonelg}) it seems likely that the latter apply also
to the ELG of the different morphological classes. However, it should
be remembered that galaxies with weaker emission lines may not have
been recognized as ELG, and those ELG that went unrecognized will have
been included in the non-ELG class. This means that, if such ELG
(i.e. with emission lines below our detection limit) do exist and if
they also have a bluer continuum than the true non-ELG, the difference
between ELG and non-ELG could be larger than shown in
Fig.~\ref{f-elg-nonelg}.

In view of the considerably smaller statistics of (in particular, the
early-type) ELG, we refrain from a detailed analysis of the strengths
of absorption and emission lines in the spectra of the various
morphological classes among the ELG. We find only one significant
effect, viz. that the $H\beta$ (486.1 nm) emission line is
$0.09\pm0.03$ mag brighter in the average spectrum of the spiral ELG
than it is in that of the S0 ELG.

\section{Discussion and conclusions}
\label{s-discuss} 

We used CCD-images to determine the morphological types of 2295
galaxies in the ENACS sample. From a comparison with morphological
types derived by Dressler (1980) for a subset of these galaxies, we
conclude that the reliability of our morphological classification is
very comparable to that of Dressler and, by implication, to that of
other expert classsifiers. With these new morphologies we could show
that the spectral classification of paper VI (based on the ENACS
spectra), allows the broad distinction between early- and late-type
galaxies to be made with a reliability of $83\pm1$\%.  However, within
the early-type class, the spectral classification can barely
distinguish ellipticals from S0 galaxies. On the contrary, late
spirals can be classified {\em from the spectrum alone} with
$72\pm5$\% reliability.

We compared the average spectral properties of galaxies without
emission lines (non-ELG) with different morphologies and luminosities,
and we reached the following conclusions.

The continuum colors, and the strength of the metal lines are strongly
and significantly correlated, with a very systematic decrease of
metallicity and a blueing of the color towards late Hubble type. In
addition, the faint galaxies in each class are generally bluer and
less metal-rich than the bright galaxies. For the colors the latter is
a confirmation of the well-established color-magnitude relation found
previously for early-type galaxies (see e.g.\ Bower et al.\ 1992).
However, our data show that this relation also extends to the spirals.
It is perhaps noteworthy that, within this very strong global trend,
the colors and metallicities of faint S0 galaxies and bright early
spirals are very similar.

Peletier et al.\ (1999) found that the bulges of nearby early spirals
have colors at their half-light radii that are very similar to those
of the ellipticals in Coma, which suggests that bulges of ellipticals
and early spirals are similar. However, early spirals have bluer
colors than ellipticals, which implies that the disk must contribute
significantly. Our observations show that galaxies of later Hubble
types and galaxies of lower luminosities indeed have bluer continua
and weaker metal lines. At the same time the average bulge fraction
decreases towards later types (see e.g.\ Kent 1985; Simien \& de
Vaucouleurs 1986; Schechter \& Dressler 1987) and towards lower
luminosities (e.g.\ Gavazzi et al.\ 2000). The spectral properties of
galaxies might therefore be mainly correlated with the bulge fraction.

The bright early spirals on average have somewhat stronger H$\delta$
absorption than the other galaxies, although in our data the effect is
at the $\sim 2\sigma$-level only. The H$\delta$ absorption-line is an
indicator for recent starformation (e.g.\ Dressler et al.\ 1999).
Observations of galaxies in clusters at $z\sim0.5$ have shown that in
many galaxies star formation was quenched after a final starburst
(e.g.\ Dressler \& Gunn 1983; Couch \& Sharples 1987), which probably
occurred when galaxies entered the cluster (Poggianti et al.\ 1999;
Ellingson et al.\ 2001). Poggianti et al.\ concluded that the
transformation from spirals to S0 galaxies occurred after the
starburst phase and on longer time scales, because the observed
post-starburst galaxies were mainly spirals instead of S0 galaxies. It
is thus possible that the transformation of early spirals into S0
galaxies is presently still going on, albeit probably at a fairly low
level.

The analysis of the properties of the galaxies with emission lines
(ELG) gives the following results.

The continuum spectrum of the ELG is bluer than that of the non-ELG, and
the blueing is, to first order, identical for galaxies of all
morphological types. This could imply that the primary source for the
line-excitation is photo-ionization

The fraction of ELG depends strongly on galaxy type: from $4\pm1$\%
for ellipticals to $59\pm4$\% for late spirals (including irregular
galaxies). This is consistent with the result obtained in paper III,
which was based on the subset of ENACS galaxies with Dressler
morphological types (about a factor of 4 smaller than the present
sample). It is also consistent with the finding by Solanes et al.\
(2001) that the fraction of gas-rich galaxies increases strongly
towards late Hubble types. According to Solanes et al.\ this indicates
that gas is removed more easily from early spirals than from late
spirals.

On the contrary, for each morphological type, the fraction of ELG
varies relatively little with projected distance from the cluster
center. This is consistent with the conclusion of Dale et al.\ (2001),
who found that the fraction of gas-rich galaxies decreases only
slightly towards the dense cluster center. Apparently, the efficiency
of the gas removal is more related to the type of galaxy than to its
position in the cluster. However, one cannot conclude from this result
that the efficiency of the process by which the gas is expelled is
{\em independent} of position within the cluster. This is because the
removal of the gas is likely to be connected to processes that affect
the galaxies in other ways, such as tidal destruction (see e.g. paper
XI) and morphological transformation (see paper IX).

\begin{acknowledgements}

The European Southern Observatory and La Palma Observatory are
acknowledged for assigning observing time for this project. We thank
Roland den Hartog and Pascal de Theije who played an important r\^ole
in the (early stages of the) imaging program, and Andrea Biviano for
helpful comments on an earlier version of the manuscript. The
observations for the imaging program on the Dutch telescope at La
Silla were made by: Jasper Arts, Jos de Bruijne, Erik Brogt, Veerle
Coupe, Peter van Dokkum, Ronnie Hoogerwerf, Magiel Janson, Dirk
Janssens, Ciska Kemper, Jaron Kurk, Arjen van der Meer, Richard Naber,
Celeste Ponsioen and Tom Voskes.

\end{acknowledgements}

\vfill
\end{document}